\documentclass{cjpsuppl}
\usepackage{graphicx}
\begin{document}
\title{Photoproduction of heavy vector meson at the LHC}
\authori{Joakim Nystrand}
\addressi{Department of Physics and Technology, University of Bergen, \\ 
All{\'e}gaten 55, N-5007 Bergen, Norway}
\authorii{}    \addressii{}
\authoriii{}   \addressiii{}
\authoriv{}    \addressiv{}
\authorv{}     \addressv{}
\authorvi{}    \addressvi{}
\headtitle{Photoproduction of Heavy Vector Mesons at the LHC}
\headauthor{J. Nystrand}
\lastevenhead{J. Nystrand: Photoproduction of heavy vector meson at the LHC}
\refnum{}
\daterec{}
\suppl{A}  \year{2005} \setcounter{page}{1}
\maketitle

\vspace{0.4cm}

\begin{abstract}
The strong electromagnetic fields associated with high energy protons and 
nuclei may lead to exclusive photoproduction of vector 
mesons in proton-proton and nucleus-nucleus collisions at the LHC. This 
paper will discuss the expected cross sections and rapidity and transverse 
momentum distributions. 
\end{abstract}

\vspace{0.4cm}


Proton-proton and ultra-peripheral nucleus-nucleus collisions at the 
Large Hadron Collider (LHC) will allow two-photon and photon-nucleon 
interactions to be studied at energies higher than at any existing 
accelerator. A photon from the electromagnetic field of one of the
projectiles may interact with the other projectile in a variety of ways. 
For a recent review of so-called ultra-peripheral collisions (UPC), 
see \cite{arnps}. This paper will deal with exclusive photoproduction 
of vector mesons \cite{prl04}. 

\begin{figure}[!t]
\begin{center}
\includegraphics[width=9cm]{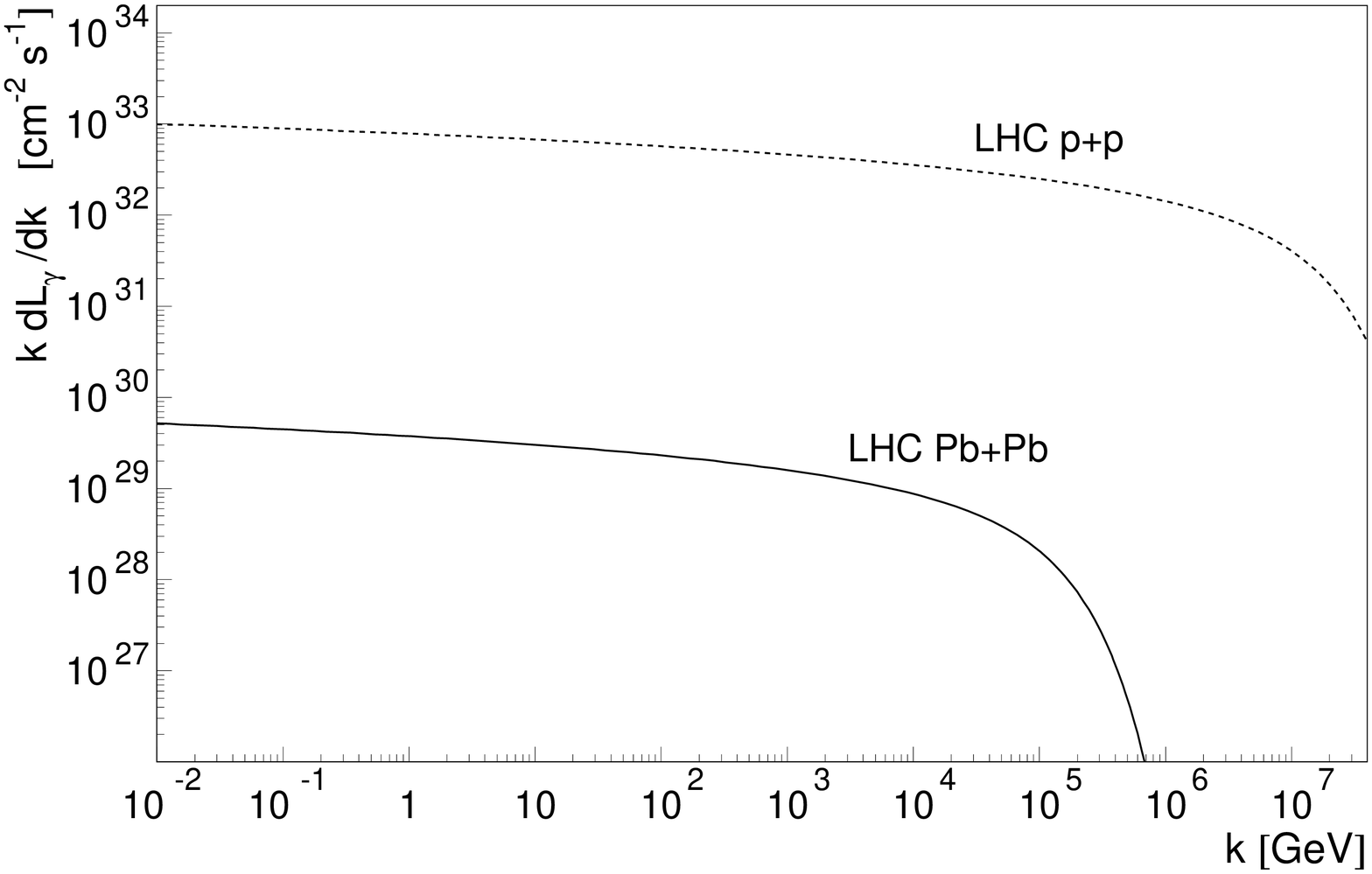}
\caption{The equivalent photon luminosity in p+p and Pb+Pb collisions at the LHC. 
$k$ is the photon energy in the rest frame of one of the projectiles. The calculations 
are for the collider luminosities ${\cal L}_{pp} = 1.0 \cdot 10^{34}$~cm$^{-2}$~s$^{-1}$ and 
${\cal L}_{PbPb} = 1.0 \cdot 10^{27}$~cm$^{-2}$~s$^{-1}$.}
\label{luminosity}
\end{center}
\end{figure}

As was first pointed out by Fermi, the effect of the electromagnetic 
field of a moving, charged particle is equivalent to that of a 
corresponding flux of photons with a certain energy spectrum. The equivalent 
photon spectrum depends on the velocity of projectile and can be calculated 
from the form factor (most appropriate for protons \cite{ppgamma}) or 
in the impact parameter representation (most appropriate for nuclei \cite{aagamma}). 
The equivalent photon spectrum in proton-proton (pp) and nucleus-nucleus (AA) collisions 
has been discussed by several authors, for details see \cite{ppgamma,aagamma} and 
references therein. 
From the photon energy spectrum, $dn_{\gamma}/dk$, the equivalent 
photon luminosity is defined by
\begin{equation}
\frac{d {\cal L}}{dk} = {\cal L}_{pp/AA} \, \frac{dn_{\gamma}}{dk} \; ,
\end{equation}
where ${\cal L}_{pp/AA}$ is the collider luminosity and $k$ is the photon 
energy. This quantity is useful for comparing the photon fluxes at different 
accelerators and for different colliding systems. The photon luminosities in 
proton-proton ($\sqrt{s} =$~14~TeV) and Pb+Pb ($\sqrt{s} =$~5.5~A~TeV) 
interactions at the LHC are shown in Fig.~\ref{luminosity}.

The photon spectrum of a nucleus with charge 
$Z$ is proportional to $Z^2$. This enhancement is, however, not enough 
to compensate for the lower beam luminosity expected at the LHC for 
Pb+Pb interactions compared with p+p interactions, as can be seen in 
Fig.~\ref{luminosity}. The photon spectrum in Pb+Pb interactions is 
furthermore cut-off at a lower photon energy owing to the larger 
minimum impact parameter. Nevertheless, nuclear beams will add to 
the physics potential of ultra-peripheral collisions. For example, 
the modification of the parton distribution functions in nuclei 
(shadowing) can be studied by comparing photoproduction on proton and 
nuclear targets . From an experimental point of view, the identification 
of exclusive events might also be easier with nuclear beams, as will be 
discussed below. 

The exclusive photoproduction of heavy vector mesons on proton and nuclear targets, 
\begin{equation}
\gamma + p \rightarrow V + p \;\;\;\;\;\; {\rm or} \;\;\;\;\;\;  
\gamma + A \rightarrow V + A \, , 
\end{equation}
has been studied over a wide energy range. 
The exclusive production of the light vector mesons $\rho$, $\omega$ and 
$\phi$ is usually described by the exchange of a soft Pomeron, and the 
cross section accordingly rises slowly with the photon-proton 
center-of-mass energy 
$W_{\gamma p}$, $\sigma \propto W_{\gamma p}^{0.22}$ \cite{Landshoff}. 
Fixed target experiments and experiments at the electron-proton collider 
HERA have found that the cross section for exclusive $J / \Psi$ production 
rises much faster with $W_{\gamma p}$. The total cross section can be 
parameterized as 
\begin{equation}
\sigma_{\gamma p \rightarrow J/ \Psi p} = 1.5 \cdot W_{\gamma p}^{0.8} \; [nb] \; ,
\end{equation}
with $W_{\gamma p}$ in GeV. This has been interpreted as evidence for the existence
of a hard Pomeron \cite{Landshoff}. In QCD based models, where the $J/ \Psi$ is produced 
via two-gluon exchange, it is seen as a consequence of the increased gluon 
density at low Bjorken-x. The $J/ \Psi$ cross section is then  
proportional to the gluon density squared, 
$\sigma \propto [g(x,Q^2)]^2$ \cite{Ryskin,Frankfurt}. 

The very limited statistics for $\Upsilon$ production does 
not allow an energy dependence to be extracted from the data. QCD based 
models predict an even more rapid increase with $W_{\gamma p}$ than for the 
$J / \Psi$. The following parameterization (for $\Upsilon(1S)$) is 
consistent with the available data 
\begin{equation}
\sigma_{\gamma p \rightarrow \Upsilon p} = 0.06 \cdot W_{\gamma p}^{1.7} \; [pb] \; ,
\end{equation}
with $W_{\gamma p}$ in GeV. The statistical and systematic errors in the experimental 
cross sections correspond to a range for the normalization constant between 0.054 and 
0.175 $[pb]$. These parameterizations for the $J / \Psi$ and the 
$\Upsilon(1S)$ are used in the calculations below. 

\begin{figure}[!t]
\begin{center}
\includegraphics[width=12cm]{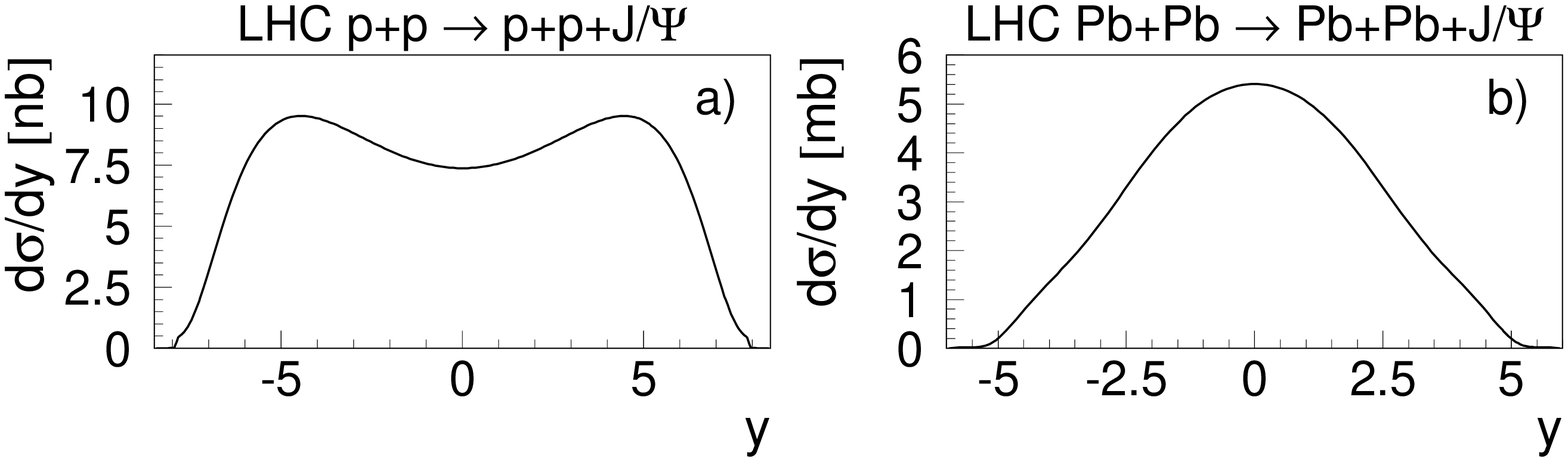}
\caption{Rapidity distributions for exclusive $J / \Psi$ production in p+p and Pb+Pb
collisions at the LHC, calculated from Eq. 5.}
\label{jpsi}
\end{center}
\end{figure}

The threshold photon energy for producing a vector meson with a certain mass, 
corresponding to $W_{\gamma p} = m_p + m_V$, is $k = m_V + m_V^2/(2 m_p)$ in the 
rest frame of the proton. This is k = 8.2 GeV and 57 GeV for the 
$J / \Psi$ and $\Upsilon$, respectively. The equivalent photon spectrum at the 
LHC extends far above these values. In fact, the maximum photon energy at the 
``knee'' in Fig.~\ref{luminosity} corresponds to a $W_{\gamma p}$ of about 
1~TeV for Pb+Pb and 10~TeV for p+p. These center-of-mass energies are much 
larger than at HERA or anywhere else. 

Heavy vector mesons can thus be produced over a wide range of photon energies at the LHC. 
There is a direct correspondence between the photon energy and the rapidity, $y$, of the 
vector meson, $y = \ln( 2 k / m_V)$, and the differential cross section for the process
$p + p \rightarrow p + p + V$ is 
\begin{equation}
\frac{d \sigma}{dy} = k \frac{dn_{\gamma}}{dk} \sigma_{\gamma p \rightarrow V p}(k) \; ,
\end{equation}
and similarly for $A + A \rightarrow A + A + V$. If the photon spectrum is known, 
the cross section $d \sigma/dy$ is thus a direct measure of the vector meson 
photoproduction cross section for a given photon energy. Both projectiles can act as 
either photon emitter or target. Interchanging the photon target and emitter 
corresponds to a reflection around $y=0$. At mid-rapidity, the photon energy is 
uniquely determined, $k = m_V/2$. For $y \neq 0$, there is a twofold ambiguity 
in $k$, depending on which projectile emitted the photon. For example, a $\Upsilon$ 
with rapidity $y=2$ can be produced by photons with energies of 35~GeV or 0.64~GeV 
in the laboratory frame. 

The photon energies for production around mid-rapidity correspond to gluon x-values 
of $(2 - 6) \times 10^{-4}$ for $J / \Psi$ production and 
$(6 - 20) \times 10^{-4}$ for $\Upsilon$ production. 
The lower number is for p+p and the higher is for Pb+Pb. 
If the twofold ambiguity in the photon energy can be resolved, i.e. 
if the contribution from the lower of the two photon energies can be estimated, 
considerably lower values of x can be reached away from mid-rapidity. 

\begin{figure}[!t]
\begin{center}
\includegraphics[width=12cm]{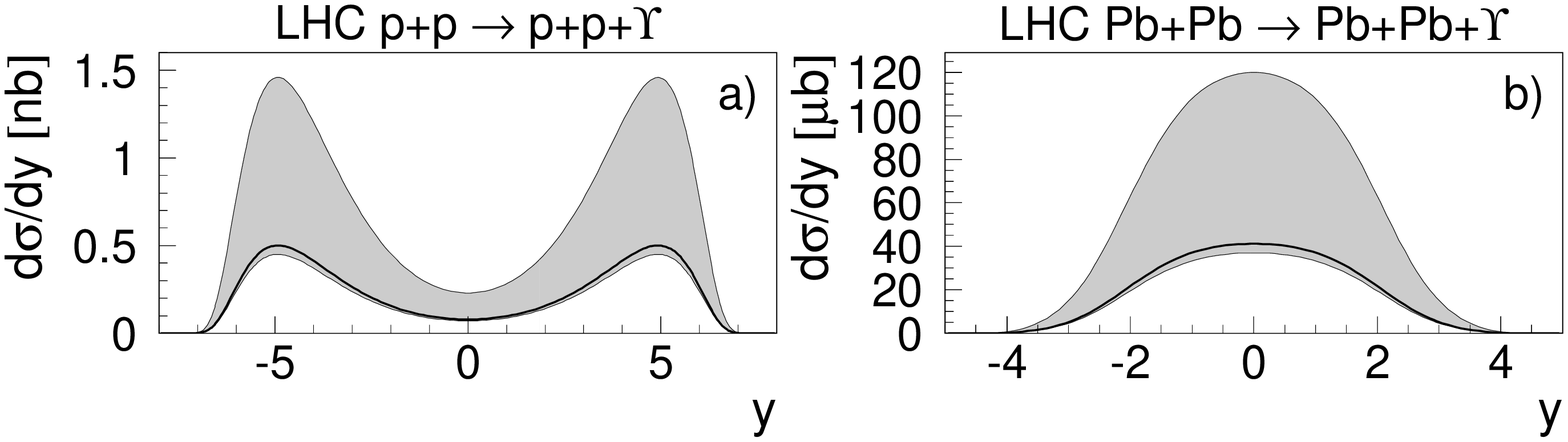}
\caption{Rapidity distributions for exclusive $\Upsilon(1S)$ production in p+p and Pb+Pb
collisions at the LHC, calculated from Eq. 5.}
\label{upsilon}
\end{center}
\end{figure}

The calculated differential cross sections as a function of rapidity are shown in 
Figs.~\ref{jpsi} and \ref{upsilon}. 
The shaded areas in Fig.~\ref{upsilon} correspond to the uncertainty in the measured 
$\sigma(\gamma p \rightarrow \Upsilon p)$. 
The integrated cross sections are listed in Table~1. For further details, see \cite{prl04}. 

The cross sections in Figs.~\ref{jpsi} and \ref{upsilon} have been calculated by adding 
the cross sections for the two photon emitter/target configurations. Under certain conditions 
the two processes may interfere \cite{prl00}. The interference will be maximal at 
mid-rapidity, where the amplitudes for the two contributions are equal because of symmetry. 
Because of the long range of the electromagnetic force and the high collision energies, 
vector mesons may be produced when the protons or nuclei are separated by several tens 
of fermi. The production will, however, always be located to within $\sim 1$~fm from the 
proton or nucleus, because of the short range of the nuclear force. 
When the spatial extension of the protons/nuclei is neglected and the produced vector 
mesons are treated as plane waves, the sum of the amplitudes 
for a given impact parameter $\vec{b}$ can be written 
\begin{equation}
| A_1 + A_2 | = 2 | A_1 |^2 [ 1 \pm cos( \vec{p_T} \cdot \vec{b} ) ] \; .
\label{eq:interference}
\end{equation}

The sign of the interference term depends on the symmetry of the system \cite{prl04}. 
It will be negative in a proton-proton 
or nucleus-nucleus collision, because of the negative parity of the 
vector meson; moving the vector meson from one production source to the other corresponds 
to a parity transformation in this case. 
In a proton-anti-proton collision, moving the vector meson 
from one production point to the other corresponds to a CP-transformation, and the sign 
of the interference term will be positive, since the vector meson has quantum numbers 
$J^{PC} = 1^{- -}$. 

\begin{table}[!b]
\caption{Cross sections for exclusive $J/ \Psi$ and $\Upsilon$ production 
 at the LHC.} \vspace{1mm}
\small
\begin{center}
\begin{tabular}{@{}lcccc}
\hline
System and energy               & $J / \Psi$ & $\Upsilon(1S)$ \\ \hline
p+p   $\sqrt{s} =$ 14 GeV       & 120 nb     & 3.5 nb         \\ 
Pb+Pb $\sqrt{s_{nn}} =$ 5.5 GeV & 32 mb      & 170 $\mu$b     \\ \hline
\end{tabular}\\
\vspace{-1mm}
\end{center}
\end{table}

\begin{figure}[thb]
\begin{center}
\includegraphics[width=10cm]{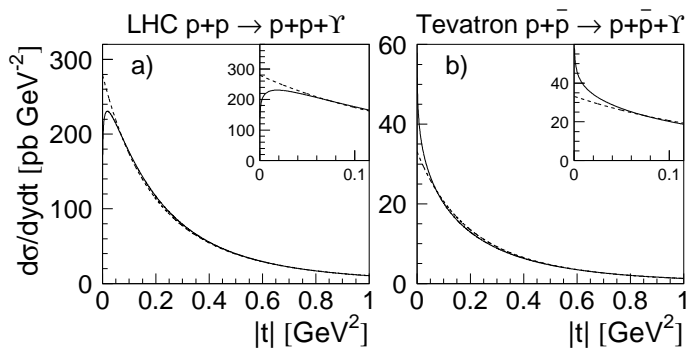}
\caption{$d \sigma / dy dt$ at $y=0$ for photoproduction of $\Upsilon$ in pp and 
$p \overline{p}$ collisions at the LHC (a) and the Tevatron (b), respectively. 
The solid curve is with and the dashed curve is without interference. The 
four-momentum transfer squared is $t \approx - p_T^2$. The inset has an expanded 
t scale.}
\label{interference}
\end{center}
\end{figure}

The vector meson transverse momentum, $p_T$, is the sum of the transverse momentum of 
the virtual photon and the momentum transfer from the target, $\sqrt{|t|}$. The latter 
is usually much larger than the former, so the vector meson transverse momentum 
distribution is determined by the form factor of the target and $p_T^2 \approx -t$. 

The interference is significant at low transverse momenta. For $p_T > 1 / b$, the 
$\cos(\vec{p_T} \cdot \vec{b} )$ term in Eq.~\ref{eq:interference} oscillates rapidly 
in the integration over impact parameter, and the interference pattern is washed out. 
For small transverse momenta, $p_T << 1 / <b>$, the scalar product 
$\vec{p_T} \cdot \vec{b}$ will be $\approx 0$ for all relevant impact parameters, and 
the interference will be noticeable. The effect on the transverse momentum distribution 
of the interference in pp collisions at the LHC is shown in Fig.~\ref{interference} a). 
For comparison, the corresponding distribution for $p \overline{p}$ interactions at 
the Tevatron is shown in Fig.~\ref{interference} b). 

The experimental feasibility of studying exclusive vector meson production in heavy ion 
reactions has been demonstrated at the Relativistic Heavy Ion Collider (RHIC) at 
Brookhaven National Laboratory. The STAR collaboration has studied the reaction 
Au+Au$\rightarrow$Au+Au+$\rho^0$ at center-of-mass energies of $\sqrt{s} =$~130 and 
200~GeV per nucleon-nucleon collision \cite{STAR02}. The identification of the 
exclusive events is achieved by their low final-state multiplicity and the very 
low transverse momentum of the produced vector meson. Preliminary results indicate 
that the interference discussed above might have been observed \cite{Klein04}. 

There are no published experimental results on exclusive vector meson production 
in $pp$ or $p \overline{p}$ interactions. The cut on low vector meson transverse 
momentum will be less efficient in pp collisions because of the different form 
factor. The identification will have to rely on the presence of so-called rapidity 
gaps, intervals in rapidity void of particles. As the discussion in \cite{prl04} shows, 
requiring one or more gaps with a total width of $\Delta y =$~3 units will probably be 
sufficient at the LHC. 

To summarize, the cross sections and rapidity and transverse momentum distributions 
for exclusive production of the heavy vector mesons $J / \Psi$ and $\Upsilon (1S)$ at
the LHC have been presented. The cross sections are large enough for these reactions 
to be studied. The center-of-mass energies at the LHC will be higher than what has 
been achieved previously. This could provide valuable information on the gluon 
distributions at low-x in protons and nuclei, and might help to resolve the nature 
of the hard Pomeron.

\bigskip

{\small I would like to acknowledge Spencer Klein (LBNL, Berkeley), 
my collaborator in this work.}

\end{document}